# Thick GEM versus thin GEM in two-phase argon avalanche detectors

A. Bondar[a], A. Buzulutskov[a]*, A. Grebenuk[a], D. Pavlyuchenko[a], Y. Tikhonov[a], A. Breskin[b]

[a]*Budker Institute of Nuclear Physics, 630090 Novosibirsk, Russia*
[b]*Weizmann Institute of Science, 76100 Rehovot, Israel*

**Abstract**

The performance of thick GEMs (THGEMs) was compared to that of thin GEMs in two-phase Ar avalanche detectors, in view of their potential application in coherent neutrino-nucleus scattering, dark-matter search and in other rare-event experiments. The detectors comprised a 1 cm thick liquid-Ar layer followed by either a double-THGEM or a triple-GEM multiplier, operated in the saturated vapor above the liquid phase. Three types of THGEMs were studied: those made of G10 and Kevlar and that with resistive electrodes (RETHGEM). Only the G10-made THGEM showed a stable performance in two-phase Ar with gains reaching 3000. Successful operation of two-phase Ar avalanche detectors with either thin- or thick-GEM multipliers was demonstrated at low detection thresholds, of 4 and 20 primary electrons respectively. Compared to the triple-GEM the double-THGEM multiplier yielded slower anode signals; this allowed applying a pulse-shape analysis to effectively reject noise signals. Noise rates of both multipliers were evaluated in two-phase Ar; with detection thresholds of 20 electrons and applying pulse-shape analysis noise levels as low as 0.007 Hz per 1 cm$^2$ of active area were reached.

## 1. Introduction

Cryogenic two-phase avalanche detectors are referred to as cryogenic detectors operated in an electron-avalanching mode in saturated vapour above the liquid phase [1,2]. Large-volume detectors of this kind have many potential applications in rare-event experiments: coherent neutrino-nucleus scattering [4], dark matter search [5,6,7], solar [8] and long baseline [9] neutrino detection.

As discussed in [2], good electron multiplication in the gas phase of two-phase detectors can be provided with Gas Electron Multipliers (GEM) [3]. Contrary to wire chambers and other "open geometry" gaseous multipliers, cascaded-GEM structures have a unique ability to operate in noble gases at high gains at room [10,11] and cryogenic temperatures [12,13,14], including in the two-phase mode in Ar [15], Kr [16] and Xe [15,17]. Particularly, high gains, reaching $10^4$, were observed in two-phase Ar avalanche detectors [15]. This permitted reaching low detection thresholds and in particular to operate two-phase avalanche detectors in a single-electron counting mode [18] and in the mode, where both ionization and scintillation signals were recorded with CsI photocathodes [19].

It should be remarked that the capability to operate with low detection threshold and low noise rate is of paramount importance for coherent neutrino-nucleus scattering and dark-matter search experiments. For example, in coherent neutrino scattering the detection threshold and noise rate should be of 1-2 electrons and of the order of $10^{-3}$ Hz per kg of the detection medium (e.g. liquid noble gas) respectively [4]; this is a real challenge in detector technology.

During the last few years the so-called thick-GEM (THGEM) concept was developed [20,21,22], motivated by the need for robust, economic, high-gain multipliers for large-volume detectors. The THGEM has about ten-fold expanded dimensions compared to the standard "thin"-GEM. It is manufactured by standard printed-circuit techniques of mechanical hole-drilling and chemical hole-rim etching in metal-clad insulating material. As compared to the "optimized GEM" [23] and LEM [24], the THGEM's etched rim assures higher multiplication. The THGEM has over an order of magnitude fewer holes than a GEM. Accordingly, it is expected to be more robust, with better resistance to discharges. In this regard, the recently proposed thick-GEM with resistive electrodes (RETHGEM) [25,26] is supposed to have even better resistance to discharges, due to the lower discharge energy. The THGEM operates at higher voltages and has lower spatial resolution compared to GEM.

* Corresponding author. Tel.: +7-383-3294833; fax: +7-383-3307163.
*E-mail address*: buzulu@inp.nsk.su

THGEM operation in pure noble gases [25,27] and at cryogenic temperatures, including in saturated vapour in Ar [25], has been the subject of recent investigations. The gains obtained in THGEMs were of the same order of magnitude as those of GEMs. On the other hand, the THGEM operation in two-phase avalanche detectors, in electron-emission from the liquid mode, had yet to be proved.

In this work, we investigated the performance of THGEM multipliers, compared to GEMs, in a two-phase Ar avalanche detector. Gain, amplitude and pulse-shape characteristics are presented for three types of THGEMs: those made of G10 and Kevlar and that of RETHGEM. An appreciable part of the work is devoted to the performance of the thin- and thick-GEM multipliers in two-phase Ar when detecting very weak signals ranging from 2 to 50 primary electrons. Noise rates in this two-phase mode were evaluated for the first time.

## 2. Experimental setup

A detailed description of the experimental setup and procedures, with regard to thin-GEM performance studies, are presented elsewhere [15,16,18,28]. Here we describe details relevant to THGEM performance in general and to THGEM and thin-GEM operation with low detection thresholds, including noise measurements.

The detector shown in Fig. 1 comprises a cathode mesh, two THGEM plates and either a G10 plate or printed circuit board (PCB); these elements, of an active area of 3×3 cm² each, were mounted in a cryogenic vacuum-insulated chamber of volume 2.5 l. The distance between the cathode and the first THGEM was 11 mm, between the THGEMs - 2 mm and between the second THGEM and the G10 plate (or the PCB) - 2 mm.

| Type | THGEM (G10) | THGEM (Kevlar) | RETHGEM |
|------|-------------|----------------|---------|
| Dielectric thickness, mm | 0.4 | 0.25 | 0.4 |
| Hole pitch, mm | 0.9 | 0.7 | 0.7 |
| Hole diameter, mm | 0.5 | 0.3 | 0.3 |
| Hole rim, mm | 0.1 | 0.1 | 0 |

Table 1. Geometrical parameters of thick-GEMs studied in this work.

Three types of THGEMs were studied: those made of G10 and Kevlar with copper electrodes, called THGEM(G10) and THGEM(Kevlar) respectively, produced for the Weizmann Institute [21] by Print Electronic Ltd., and that made of G10 with resistive electrodes using screen printing technique, called RETHGEM, produced at the CERN workshop [26].

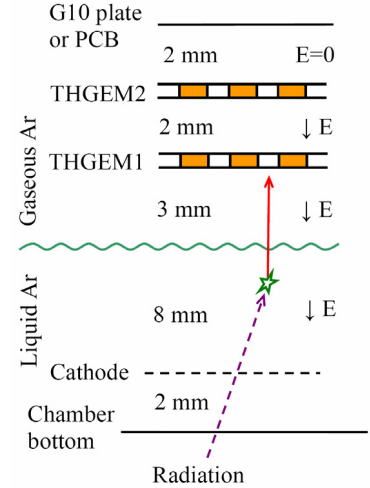

Fig. 1. Schematic view of the experimental setup to study THGEM performance in two-phase Ar avalanche detector in electron emission mode, i.e. at standard direction of the electric field within liquid Ar.

Their geometrical parameters are presented in Table 1. Note that in contrast to the THGEM(G10) and THGEM(Kevlar) the RETHGEM had no hole rims.

The thin-GEMs were produced by the CERN workshop and had the "standard" parameters: 50 and 5 μm thickness of kapton and copper clad respectively, 140 μm hole pitch, 70 and 55 μm hole diameter in metal and kapton respectively, 28×28 mm² active area. In the measurements with thin-GEMs, three cascaded GEM foils were mounted in the cryogenic chamber.

In the two-phase mode the detector was operated in Ar in equilibrium state close to the triple point, at a temperature of 84 K and vapour pressure of 0.70 atm. At these conditions, the thickness of the liquid condensate at the chamber bottom was approximately 10 mm, corresponding to the active mass of liquid Ar of about 10 g, and the distance between the liquid surface and the first THGEM - 3 mm. The electron life-time in liquid Ar before attachment was larger than 20 μs, corresponding to a drift path of 5 cm at an electric field in the liquid Ar of 1.5 kV/cm [28]. Usually it took 3 hours to reach the equilibrium state after the beginning of cooling. Some measurements in the two-phase mode were done also during the warming-up procedure, when the cooling was stopped and the pressure and the temperature gradually increased. This procedure took typically half an hour, until reaching the point where the liquid fully evaporated.

The cathode and THGEM electrodes were biased through a resistive high-voltage divider placed outside the cryostat (Fig. 2). Each electrode was connected to the divider using high-voltage feedthroughs and 1 m long wires. The anode signal was read out from the divider using a charge-sensitive preamplifier, with a 10 ns rise time and sensitivity of 0.5 V/pC; it was followed by a research amplifier (ORTEC 570) with an amplification factor of 20 and shaping times of either 0.5 or 10 μs; the former was used for pulse-shape analysis and the latter for gain measurements. Both shaping-times were also used for measuring amplitude spectra. The signals were analyzed with a TDS5032B digital oscilloscope; it allowed to store in the memory up to 5000 waveforms for the subsequent offline analysis.

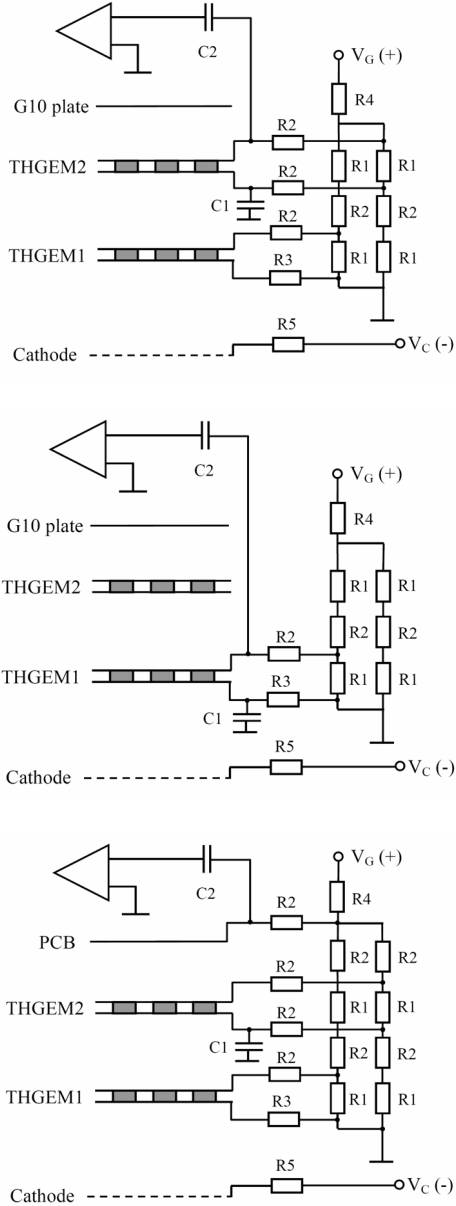

Fig. 2. Electrical schemes of the THGEM multiplier, showing the high-voltage divider and readout electronics in "2THGEM" (top), "1THGEM" (middle) and "2THGEM+PCB" (bottom) operation modes. The resistance values of R1, R2, R3, R4 and R5 are 8.6, 4.3, 10, 3.3 and 4.8 MOhm, respectively.

The single- and double-THGEM multipliers were operated in "1THGEM" and "2THGEM" modes respectively (see Fig. 2). In these modes the anode signal was read out from the last electrode of the first and second THGEM respectively. In the 1THGEM mode the second THGEM was under floating potential. In the 2THGEM mode either the G10 plate was under floating potential (Fig. 1) or the PCB was kept at more negative potential than that of the second THGEM (Fig. 3). Such choices of potentials guaranteed that the field-lines are terminated at the readout electrode, providing full collection of the avalanche electrons. In one specific case the double-THGEM multiplier was operated in a "2THGEM+PCB" mode, where the anode signal was read out from the PCB, as shown in Fig. 2.

In the current work the thin-GEM multiplier was operated only in a "3GEM" mode.

The detector was irradiated from outside by either a pulsed X-ray tube or $^{241}$Am X-ray source; the latter providing, among others, a 60 keV X-ray line. When using the pulsed X-ray tube, the measurements were done in a triggered mode, the trigger signal being provided by the X-ray generator.

The gain of the THGEM multiplier was measured with the pulsed X-rays with the amplifier's shaping time of 10 µs, similar to that in our previous works [15]: the gain is defined as the pulse-height of the avalanche (anode) signal divided by that of the calibration signal. The latter was recorded at the first electrode of the first THGEM, with no high voltage applied to the divider. At higher gains, the gain-voltage dependence was measured using the relative peak position in the pulse-height spectrum, induced by 60 keV X-rays. The two sets of measurements were joined, providing the absolute gain values. The maximum gain was defined at the onset of discharges.

To measure amplitude spectra, two methods were employed. In the first, the anode signal was electronically integrated with amplifier shaping time of 10 µs and the pulse-height of the signal was recorded. In the second method, the anode signal was differentiated with a shaping time of 0.5 µs and the signal integration was done off-line by calculating the pulse area.

The amplitude of the anode signals in each of investigated multipliers is proportional to the initial charge (prior to multiplication) arriving to the first THGEM. In the following this charge is called "primary electrons". To express the amplitude of the anode signal in terms of primary electrons ($A$), the measured amplitude ($S$) should be divided by the calibration coefficient of the electronic circuit ($C$), obtained

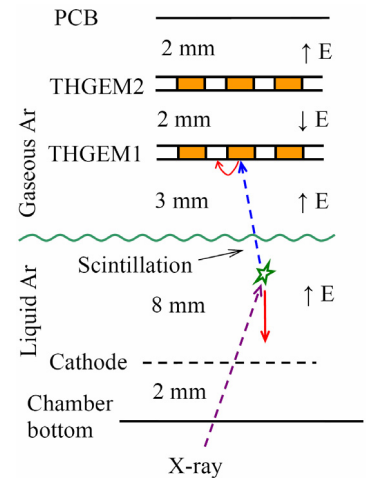

Fig. 3. Schematic view of the experimental setup to study THGEM performance in two-phase Ar avalanche detector in non-emission mode, i.e. with reversed or close to zero electric field within liquid Ar and in the drift region between the liquid and first THGEM.

with a pulse generator, and by the multiplier's gain ($G$). For example, in the first method it is:

$$A[e] = \frac{S[V]}{C[V/e]G}$$

In addition to operation in electron-emission mode, illustrated in Fig. 1, the detector was also operated in non-emission mode, illustrated in Fig. 3. The latter permitted to study the GEM- and THGEM-multiplier performance for detecting weak signals, induced by 2 to 50 primary electrons, similarly to that studied elsewhere for thin-GEMs [18]. This mode was also used to measure the noise of each of the multipliers investigated. The "emission" (electron emission from liquid) and "non-emission" modes of operation respectively corresponded to the "normal" and "reversed" (or close to zero) electric field across the liquid Ar (compare Fig.1 to Fig. 3). In non-emission mode the signal was induced by photoelectrons from the first electrode of the first THGEM, originating from UV-scintillations in the liquid (Fig. 3).

## 3. Gain and amplitude characteristics

Rather high gains, exceeding $10^3$ and $10^4$ in 1THGEM and 2THGEM mode, respectively, were observed in THGEM multipliers made of G10 and Kevlar in gaseous Ar at room temperature; their gain-voltage characteristics at 1 and 1.9 atm, at a drift field of 0.3 kV/cm, are presented in Figs. 4 and 5. Note that the onset of dischargers was not reached here. Such high gains are in accordance with previous measurements [25,27].

The maximum attainable gain of the double-RETHGEM in Ar at room temperature was only a few thousands (Fig. 5); it is in accordance with the gain reported elsewhere for RETHGEMs produced by the screen-printing technique [26].

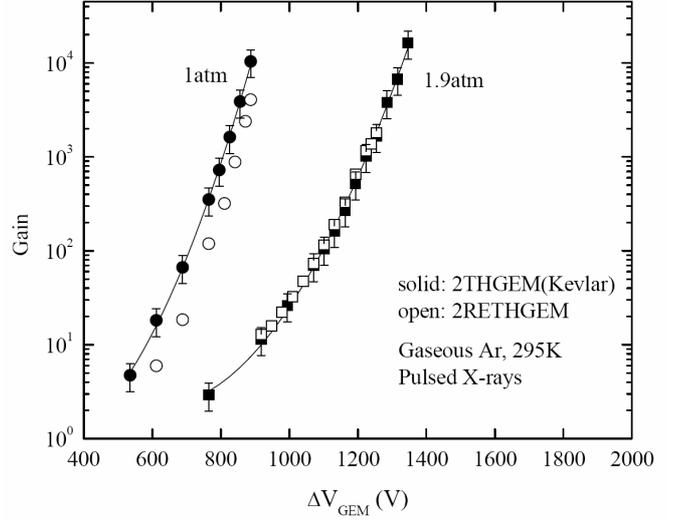

Fig. 5. Gain-voltage characteristics of double-THGEM(Kevlar) and double-RETHGEM multipliers in gaseous Ar at room temperature at 1 and 1.9 atm, measured using pulsed X-rays. The maximum gains for RETHGEM were limited by discharges, while for THGEM(Kevlar) the discharge limits were not reached.

The lower gains obtained for RETGEM (compared to THGEM) are most probably due to the absence of hole-rims which are known to provide a substantial increase of the maximum gain [21].

Gain-voltage characteristics of THGEM multipliers in two-phase Ar avalanche detectors, in electron-emission mode, are presented in Figs. 6 and 7. Fig. 6 shows the characteristics of the double-THGEM(G10) and double-THGEM(Kevlar) multipliers in comparison with a typical characteristic of a triple-GEM. Rather high gains were reached: 3000 and 6000 in the double-THGEM(G10) and in the double-THGEM(Kevlar) respectively, compared to $10^4$ for that of the triple-GEM.

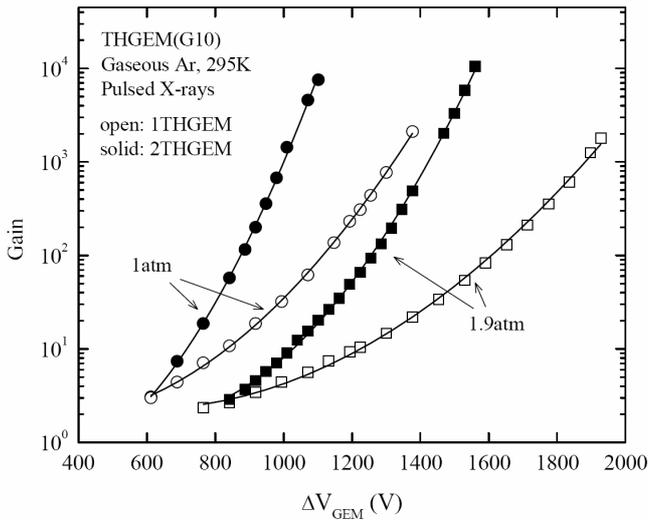

Fig. 4. Gain-voltage characteristics of single- and double-THGEM(G10) multipliers in gaseous Ar at room temperature at 1 and 1.9 atm, measured using pulsed X-rays. Gains as a function of the voltage across each THGEM are shown in 1THGEM and 2THGEM modes. The discharge limits were not reached.

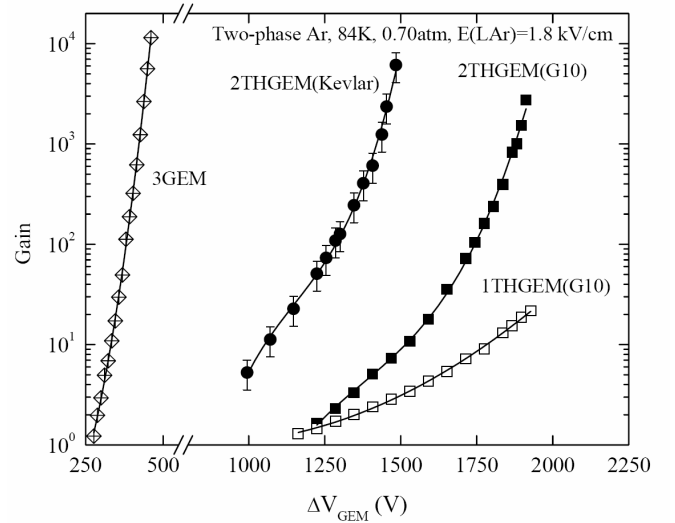

Fig. 6. Gain-voltage characteristics of single- and double-THGEM(G10) and double-THGEM(Kevlar) multipliers in two-phase Ar in electron emission mode at 84 K and 0.70 atm, measured with pulsed X-rays and $^{241}$Am X-rays. For comparison, the characteristic of a triple-GEM is shown. The maximum gains were limited by discharges (except of that in the single-THGEM). The electric field within liquid Ar E(LAr)=1.8 kV/cm.

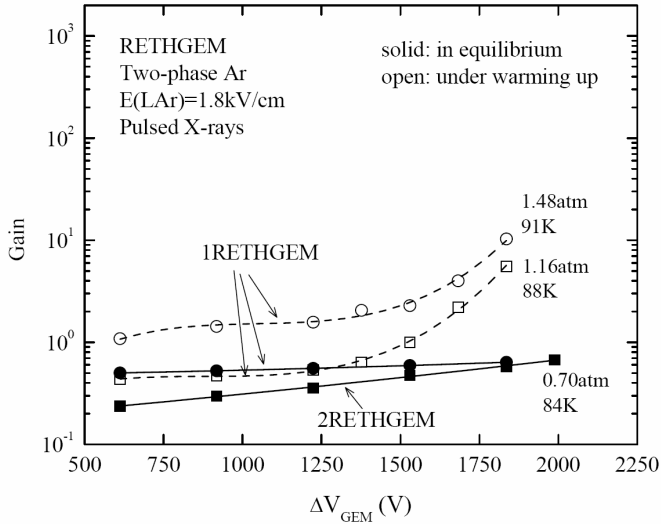

Fig. 7. Gain-voltage characteristics of single- and double-RETHGEM multipliers in two-phase Ar in electron emission mode in equilibrium and under warming-up, measured with pulsed X-rays. The maximum gains were limited by discharges. The pressures and temperatures are indicated in the figure.

It should be remarked that the THGEM's gain increases with voltage substantially faster than an exponential function, in contrast to the exponential rise in thin-GEMs. This fact is probably related to the pulse-width broadening with the avalanche charge, not observed for thin-GEMs; it will be discussed in the next section.

Compared to its performance at room temperature, the present RETHGEM did not operate properly in the two-phase mode: no multiplication was observed neither in single- nor in double-RETHGEM when the two-phase Ar detector was in an equilibrium state (see Fig. 7). Only under warming-up conditions, when there was a substantial temperature gradient inside the chamber, the RETHGEM could weakly multiply electrons in two-phase Ar (Fig. 7). An unusual gain dependence on pressure should be noted at these conditions: at a given voltage the gain is higher for higher pressure. Such a behavior could be explained by liquid Ar condensation within the RETHGEM holes in the equilibrium state and by its evaporation from the holes under the warming-up procedure.

The difference in gain between RETHGEM and the other multipliers is not yet understood and requires further investigations. Possible reasons might be a higher wetting properties of the RETHGEM electrode material, defects in the resistive layer appearing at low temperatures, etc.

The pulse-height resolution of the two-phase Ar avalanche detector, in electron-emission mode, using the double THGEM(G10) multiplier, is illustrated in Fig. 8: the amplitude spectra obtained using the two methods described in the previous section are presented. Both methods yielded similar resolutions: 18% RMS at 60 keV; it is practically the same to that of the two-phase Ar avalanche detector using a triple-GEM [15,19]. Note that for 60 keV deposited in the liquid, the anode signal corresponded to about 1000 primary electrons

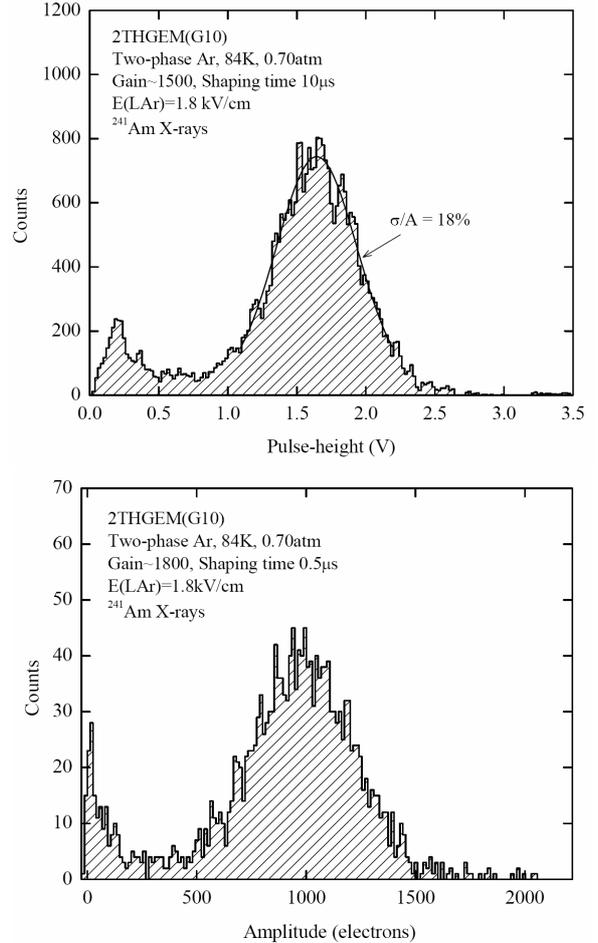

Fig. 8. Amplitude spectra of anode signals of a double-THGEM(G10) multiplier in two-phase Ar, in electron emission mode, induced by $^{241}$Am X-rays. Top: pulse-height spectrum at amplifier shaping time of 10 μs and detector gain of 1500. Bottom: spectrum of pulse-area at amplifier shaping time of 0.5 μs and detector gain of 1800; the abscissa represents the initial charge prior to multiplication, i.e. the number of primary electrons (see Fig. 8, bottom).

It was found that THGEMs made of Kevlar exhibited a rather strong charging-up effects, both at room temperature and in the two-phase mode (see Fig. 9): after supplying the voltage the anode pulse-height substantially increased with time within half an hour. On the other hand, no charging-up effects were observed for THGEMs made of G10 (Fig. 9) and for RETHGEMs. The charging-up effect resulted in a poor amplitude resolution of the two-phase Ar avalanche detector when using the THGEM(Kevlar) multiplier, as seen in Fig. 10. The degraded spectrum could originate from gain non-uniformity due to a variation of charging-up time constants from hole to hole. Note that the data presented in Figs. 5, 6 and 10 for the THGEM(Kevlar) multiplier were obtained in half an hour after applying the high voltage, to minimize uncertainties due to the charging-up effect.

Thus we may conclude that from three types of THGEMs studied in the present work, THGEMs made of G10 showed the most satisfactory performance in the two-phase Ar avalanche detector. Accordingly, in the next sections the pulse-shape and noise characteristics will be considered for THGEMs made of G10, in comparison with those of thin-GEMs.

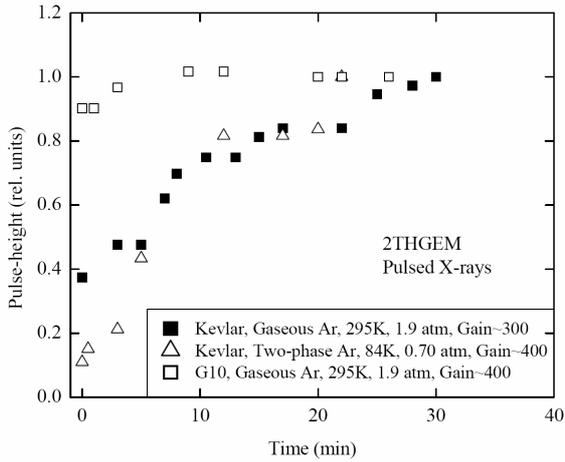

Fig. 9. Stability test. Pulse-height variation with time (after HV application) of anode signals of a double-THGEM(Kevlar) multiplier in gaseous Ar at room temperature and in two-phase Ar, induced by pulsed X-rays. The behavior of a double-THGEM(G10) multiplier is shown for comparison. The conditions are indicated in the figure; the gains were measured 30 min after HV application.

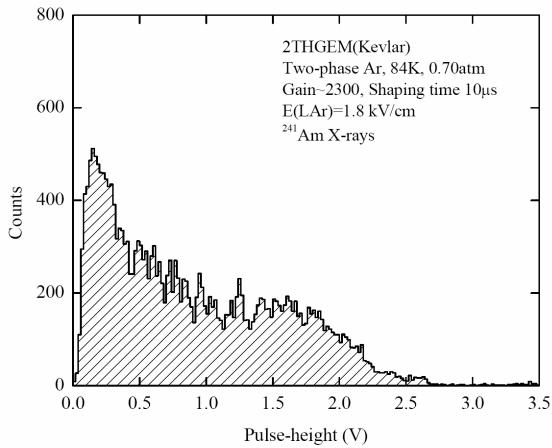

Fig. 10. Pulse-height spectrum of anode signals of a double-THGEM(Kevlar) multiplier in two-phase Ar, in electron emission mode; signals were induced by $^{241}$Am X-rays, at a detector gain of 2300 and amplifier shaping time of 10 μs.

## 4. Pulse-shape characteristics

In this section we present the pulse-shape analysis of anode signals of the double-THGEM(G10) in two-phase Ar, in comparison with those of the triple-GEM. In order to study the time structure of the signals, the analysis was done at the lower amplifier shaping time, i.e. at 0.5 μs.

A typical anode signal of the triple-GEM in two-phase Ar, induced by a 60 keV X-ray, is shown in Fig.11. Since the anode signal in a thin-GEM is fast, one can study the time structure of electron emission through the liquid-gas interface. In particular, it was observed that the signal had a slow emission component, with a width of about 8 μs (FWHM), in addition to the fast emission component. The slow emission component in two-phase Ar was observed earlier [29,30]; its physical nature was explained by thermionic emission of electrons in the frame of electric-field-enhanced Schottky model [30].

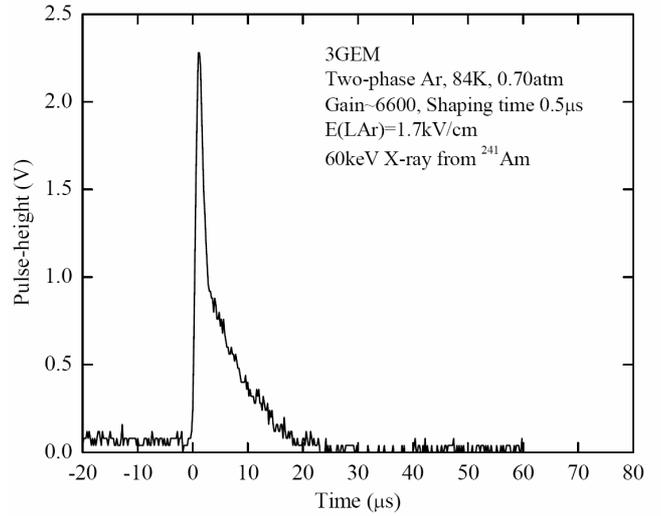

Fig. 11. Typical anode signal of a triple-GEM operated in two-phase Ar in electron-emission mode, induced by 60 keV X-ray from $^{241}$Am at a detector gain of 6600. Fast and slow emission components are distinctly seen.

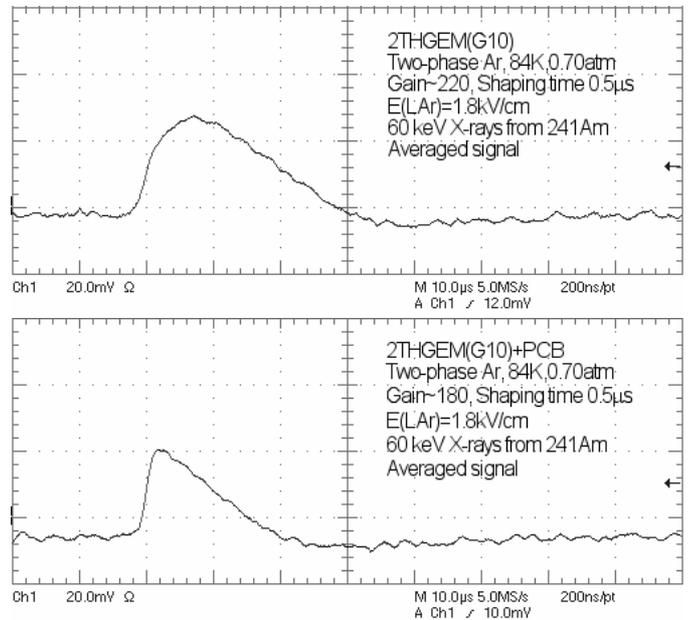

Fig. 12. Typical anode signals of a double-THGEM(G10) multiplier operated in 2THGEM (top) and 2THGEM+PCB (bottom) modes in two-phase Ar, in electron-emission mode. Signals were induced by 60 keV X-rays from $^{241}$Am, at respective gains of 220 and 180. The signals were averaged over 8 events.

A typical anode signal of the double-THGEM(G10) in two-phase Ar at a moderate gain (~200), induced by a 60 keV X-ray, is shown in Fig.12. The fast component is not seen, the pulse rise is "rounded" at a microsecond scale and the signal width (~18 μs) is significantly larger than that of the triple-GEM.

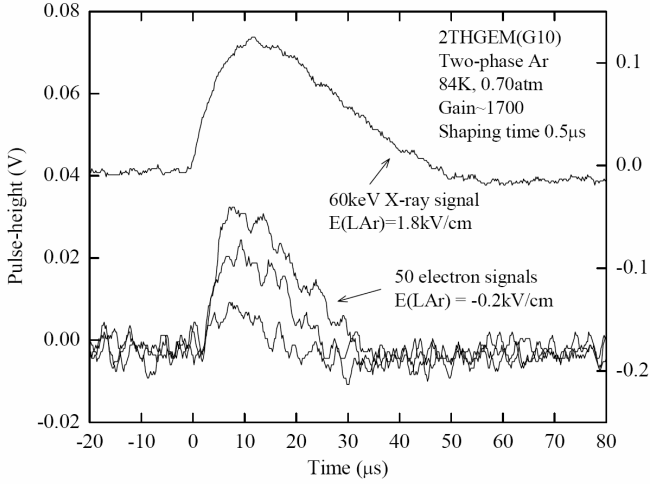

Fig. 13. Typical anode signals of a double-THGEM(G10) (2THGEM mode) operated in two-phase Ar; pulses were induced by 60 keV X-ray from $^{241}$Am (inducing ~1000 primary electrons) in electron-emission mode (top, right scale), and by pulsed X-rays producing 50 primary electrons on the average, in non-emission mode (bottom, left scale). Detector gain is 1700.

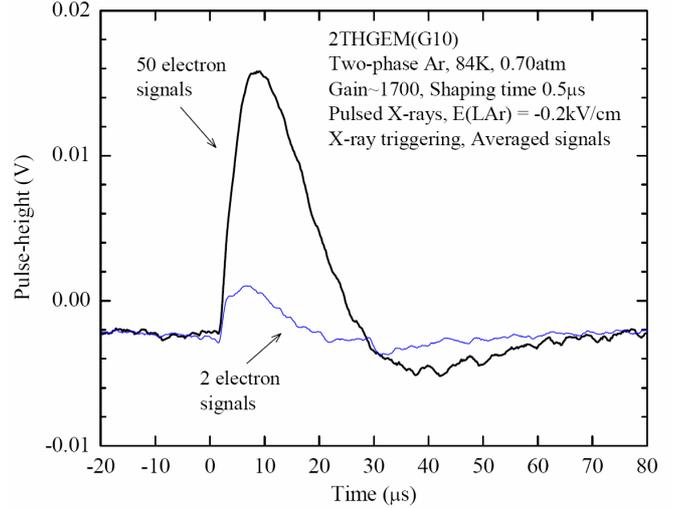

Fig. 14. Anode signals recorded in triggered mode in a double-THGEM(G10) (2THGEM mode) in two-phase Ar in non-emission mode. The signals were induced by pulsed X-rays yielding 50 and 2 primary electrons on the average, at a detector gain of 1700. The signals were averaged over several tens events.

These features could be explained by the fact that the anode signal of the THGEM is inherently slow. Its major part is induced by ions drifting between the electrodes of the second THGEM; their drift-time turned out to be larger in THGEMs, in contrast to thin-GEMs, essentially due to larger drift path within the former.

Our hypothesis was confirmed by analyzing the pulse-shape in the 2THGEM+PCB mode. In this configuration the anode signal, induced by electrons, is not sensitive to the drifting ions. One can see in Fig. 12 that the signal is not rounded, unlike that in the 2THGEM mode; it has a "triangular" shape with a width of 8 μs, like that in the triple-GEM. That means that the signal width, in this case, is defined by the slow component of the electron emission from the liquid. On the other hand, the fast emission component was not pronounced in this case, due to a relatively large rise-time of the signal. The latter resulted from the lower time-resolution of the THGEM compared to that of the thin-GEM.

An interesting observation in the THGEM was the pulse-width broadening with gain and initial charge. This can be seen from Fig. 13 showing a typical anode signal of the double-THGEM(G10) in two-phase Ar, induced by a 60 keV X-ray at a gain of about 1700. Note that this signal was induced by a rather large initial charge, of about 1000 electrons (see Fig. 8, bottom); the width of the signal is of ~25 μs.

This should be compared to weaker signals at the same gain, induced by 50 and 2 primary electrons on the average (Figs. 13 and 14); the technique of obtaining such weak signals is described in the next section. Their pulse-widths are much smaller, of 14 and 10 μs respectively. Nevertheless, the pulse-shape of these signals is generally the same as that of signals induced by larger initial charges; they have a rounded pulse-rise on a microsecond scale. It should be emphasized that in this non-emission operation mode the pulse-shape is defined essentially by the electron avalanche and ion drift processes, and not by electron emission through the liquid-gas interface.

The data on the pulse-widths measured in the THGEM(G10) multiplier are summarized in Table 2. They indicate that the pulse-width is a function of the product of gain and the initial charge, i.e. of the total avalanche charge. The mechanism of pulse-width broadening with the avalanche charge is not yet understood. We can only speculate that it may result from space-charge and ion-clustering effects.

It will be shown in the following that the characteristic pulse-shape of the THGEM's anode avalanche signal could be used for effective noise rejection when detecting weak signals.

| Mode | 2THGEM, Non-emission | 2THGEM, Non-emission | 2THGEM, Emission | 2THGEM, Emission | 2THGEM+PCB, Emission |
|---|---|---|---|---|---|
| Gain | 1700 | 1700 | 220 | 1700 | 180 |
| Initial charge , e | 2 | 50 | 1000 | 1000 | 1000 |
| Avalanche charge G×Q, e | $3.4 \times 10^3$ | $8.5 \times 10^4$ | $2.2 \times 10^5$ | $1.7 \times 10^6$ | $1.8 \times 10^5$ |
| Pulse width, μs | 10 | 14 | 18 | 25 | 8 |

Table 2. Pulse widths (FWHM) of anode avalanche signals of the double THGEM(G10) multiplier in two-phase Ar in "emission" (Figs. 12, 13) and "non-emission" (Fig. 14) operation modes and at different gains (G) and initial charges prior to multiplication (Q).

## 5. Detection of weak signals

The possibility to detect weak signals in two-phase avalanche detectors, in the range of 1-100 initial electrons, is the key factor in coherent neutrino scattering and dark-matter search experiments [4,5,6,7]. In order to study the THGEM performance with weak signals, we developed a technique of obtaining signals with small amplitudes, in the range of 2-50 primary electrons; it is similar to that presented in [18], where it permitted reaching single-electron counting in triple-GEM multipliers. A two-phase detector was irradiated with a pulsed X-ray tube, with a reversed electric field across the liquid Ar. This resulted in the suppression of the primary ionization signal; only the photoelectric signal from the first THGEM electrode facing the liquid, induced by primary UV-scintillation in liquid Ar, was recorded (see Fig. 3).

By decreasing the X-ray intensity per pulse one could reduce the number of photoelectrons emitted from the face of the first THGEM, down to few-electrons levels. It was also observed [18] that the amplitude spectra of GEM multipliers in two-phase Ar were exponential, $N\sim exp(-A/<q>)$, with the curve's slope $<q>$ characterizing the average charge recorded. In particular, in single-electron counting mode this charge is equal to one electron on average, if the amplitude is expressed in primary electrons.

Fig. 15 shows the amplitude spectrum of the anode (avalanche) signals in the double-THGEM operating in two-phase Ar when detecting weak signals using the technique described above. The electronic noise spectrum, obtained by suppressing the X-ray radiation with a tungsten screen, is also shown; it is characterized by the RMS deviation defined as Equivalent Noise Charge (ENC), if the amplitude is expressed in charge units at the preamplifier's input. For the electronic noise spectrum in Fig. 15, ENC=4000 e.

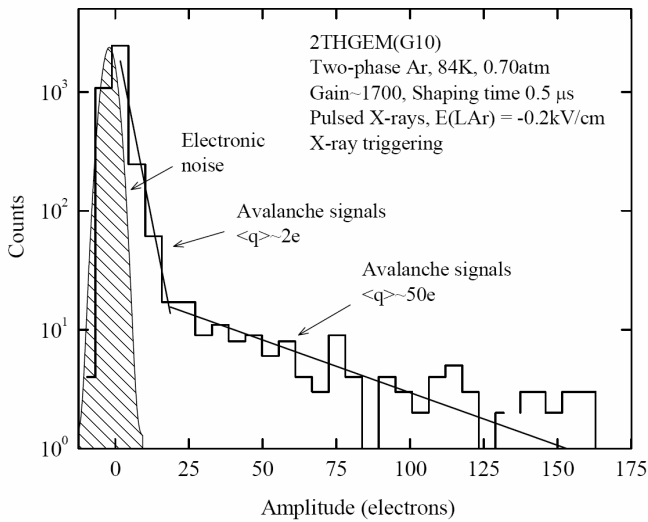

Fig. 15. Pulse-area spectrum of anode (avalanche) signals recorded in triggered mode of a 2THGEM(G10) multiplier in two-phase Ar, in non-emission mode, induced by pulsed X-rays. Amplifier shaping time is 0.5 μs and detector gain is 1700. The electronic noise spectrum is also shown. The abscissa represents the initial charge prior to multiplication.

In the present work we were not able to effectively operate the double-THGEM in two-phase Ar in single-electron counting mode due to lower gain and larger electronic noise as compared to [18]; nevertheless, the operation in the "2 electron" counting mode was shown to be possible. This is seen in Fig. 15: the avalanche-signal spectrum is separated from that of electronic noise; the average charge of the former is ~ 2 primary electrons. It is also seen that in addition to the main, low-amplitude component, the spectrum has also a high-amplitude tail; the latter was presumably induced by high-energy X-rays. The amplitude of the latter component is about 50 primary electrons on the average. The selection of either the low-amplitude or high-amplitude components in the amplitude spectrum yielded the "2 electron" or "50 electron" signals studied in the previous section.

In the previous paragraphs we considered the operation with an external X-ray trigger, with the purpose of understanding the properties and detectability of weak signals. On the other hand, in dark-matter search and coherent neutrino scattering experiments the detector should operate in self-triggered mode, with the lowest possible detection threshold above the electronic noise. In the following we will therefore investigate the operation in self-triggered mode with the purpose of defining a minimum detection threshold of two-phase avalanche detectors and studying their noise characteristics.

In the present work the minimum detection threshold (above electronic noise) of the double-THGEM multiplier in two-phase Ar was found to be 20 primary electrons with a 0.5 μs shaping time and avalanche gain of 1700; this can also be derived from Fig. 15. In electron emission mode, this amplitude would correspond to a threshold energy of 1 keV deposited in liquid Ar, if to take into account that the 60 keV X-rays induce ~1000 primary electrons.

For comparison, the minimum detection threshold of the triple-GEM multiplier in two-phase Ar was determined with a 10 μs shaping time, since the faster pulses did not allow for pulse-shape analysis anyway. At a gain of 3700 the threshold was of 4 primary electrons. This value is lower than that of the double-THGEM due to the lower noise level (with ENC=1000 e) and the higher gain. It is equivalent to an energy threshold in electron emission mode of ~0.2 keV.

## 6. Noise characteristics

To further characterize the performance of the two-phase Ar avalanche detector at low detection threshold, its noise characteristics (spectra and rates) should be studied. These were measured with the double-THGEM and triple-GEM multipliers at their minimum detection thresholds, i.e. at 20 and 4 primary electrons respectively: see Figs. 16 and 17 respectively. Note that here the matter concerns the "physical" noise, i.e. that generated by the two-phase avalanche detector itself due to cosmic rays, radioactivity, leakage currents, discharges, etc; the electronic noise was below the detection threshold.

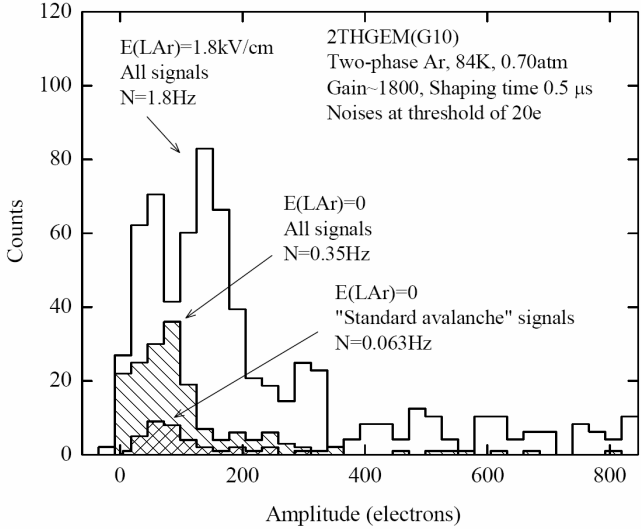

Fig. 16. Pulse-area spectra of noise (anode) signals recorded in self-triggered mode of a 2THGEM(G10) multiplier in two-phase Ar at detection threshold of 20 primary electrons. Amplifier shaping time is 0.5 µs; detector gain is 1800; measuring time is 10 min. Unshaded: electron-emission mode; shaded: non-emission mode for all signals and for signals having "standard avalanche" shapes. The abscissa represents the initial charge prior to multiplication. The appropriate noise-rates (N) are indicated.

Comparing noise rates and spectra of the double-THGEM multiplier in electron-emission and non-emission modes (Fig. 16), one may conclude that at a threshold of 20 primary electrons the main source of noise originates from the liquid; the noise rate was of the order of 1 Hz, with amplitudes exceeding 100 primary electrons. This rate should be compared to 0.2 Hz expected from cosmic rays traversing the active area and to ~ 3 Hz, when they cross the rest of the liquid.

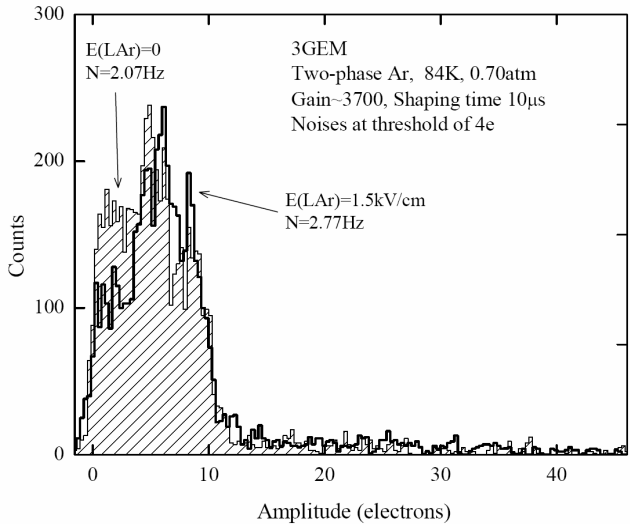

Fig. 17. Pulse-height spectra of noise (anode) signals recorded in self-triggered mode of a triple-GEM in two-phase Ar at detection threshold of 4 primary electrons, in electron emission and non-emission modes. Amplifier shaping time is 10 µs; detector gain is 3700; measuring time is 10 min. The abscissa represents the initial charge prior to multiplication. The appropriate noise-rates (N) are indicated.

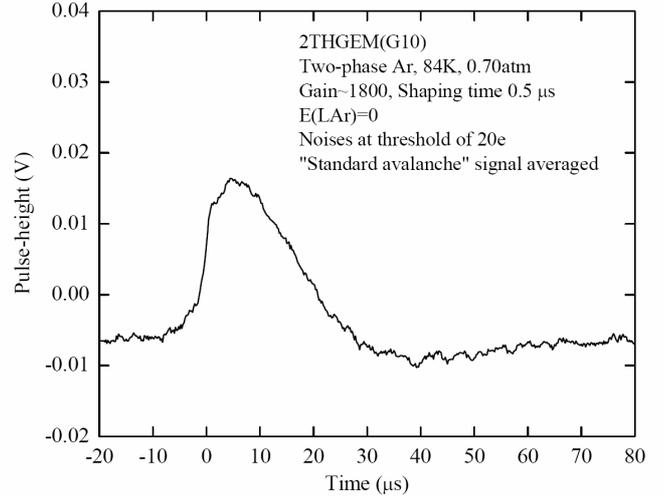

Fig. 18. Noise (anode) signals of a 2THGEM(G10) multiplier in two-phase Ar in non-emission mode, having a "standard avalanche" pulse shape. The gain is 1800 and the detection threshold is 20 primary electrons. The signals were averaged over a period of 10 min.

In the triple-GEM multiplier operating at a detection threshold of 4 primary electrons (Fig. 17), the difference in noise rates between electron emission and non-emission modes was approximately the same as that in the double-THGEM, confirming the statement that the liquid volume is a major source of characteristic noise, with ~1 Hz rate and with relatively large amplitudes. On the other hand, one can see that here the noise-rate in non-emission mode is still high, of about 2 Hz; most signals have small amplitudes, below 10 primary electrons. That means that in this case the main noise source is determined by the GEM multiplier itself; quite possible, this is due to the lower detection threshold.

In non-emission mode the noise-rate of the double-THGEM multiplier was still ~0.4 Hz (Fig. 16). We found that the effective means to further reduce the noise rate was to apply pulse-shape analysis. In particular, it was observed that all noise signals could be classified in 4 types: see Figs. 18 and 19. The first one had a "standard avalanche" pulse shape (Fig. 18) similar to that observed when detecting weak avalanche signals induced by 50 primary electrons (Fig. 14). In particular, the rounded pulse rise should be noted.

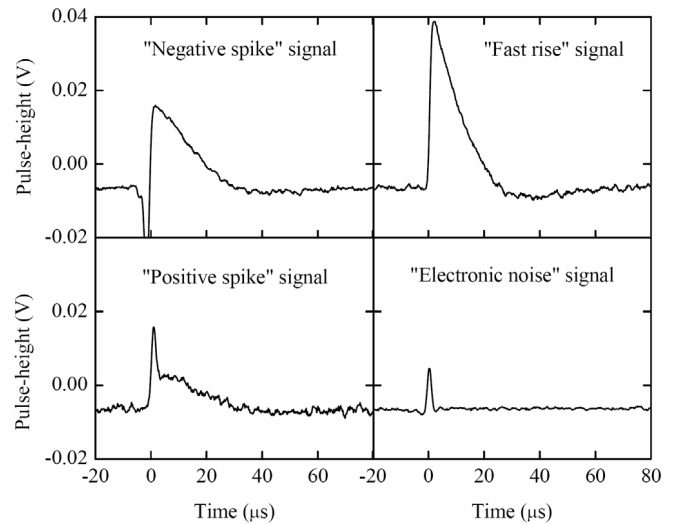

Fig. 19. Noise (anode) signals of a 2THGEM(G10) multiplier in two-phase Ar in non-emission mode, having different types of "noise-like" pulse shapes. The measurement conditions are the same as that of Fig. 18, except of the "electronic noise" signal type where the detection threshold was decreased. The signals were averaged over a period of 10 min.

The other three types of signals had different, "noise-like" pulse-shapes (Fig. 19): those with negative and positive spikes and others with a fast rise. These types of signals have low probability of being produced by particle-induced electron avalanches and thus could be rejected.

For completeness, the "electronic noise" signal, obtained at a lower detection threshold, is also shown: it is much faster than the other signals and therefore can easily be recognized and rejected.

Accordingly, only the signals having the "standard avalanche" shape were taken into consideration. This resulted in a 5-fold reduction of the double-THGEM's noise rate, to a value as low as ~0.06 Hz (Fig. 16); it corresponds to one event per 16 sec or 0.007 Hz per 1 cm$^2$ of the detector's active area. This is a rather low noise rate, in particular compared to those of photomultiplier tubes being used in two-phase detectors for dark matter search experiments [5,7].

Further studies are needed in order to decrease the noise-rate of two-phase avalanche detectors and THGEM multipliers, preferably performed with lower detection thresholds and at cleaner background environment. It should be noted that there is an ongoing R&D on THGEM production from radio-clean Cirlex (polyimide) printed-circuit boards [31].

## 7. Conclusions

The performance of two-phase Ar avalanche detectors with thick-GEMs (THGEMs) multipliers was compared to that with thin-GEMs, in view of their potential applicability for the detection of rare events. The detectors comprised a 1 cm thick liquid Ar layer and either a double-THGEM or a triple-GEM, operated in the saturated Ar vapor above the liquid phase.

Three types of THGEMs were studied in two-phase Ar: those made of G10 and Kevlar and that with resistive electrodes (RETHGEM). The G10-made double-THGEM exhibited stable operation in two-phase Ar with gains reaching 3000 and pulse-height resolution of 18% for 60 keV X-rays. The Kevlar-made double-THGEM and the double-RETHGEM exhibited operation instabilities and very low gains, respectively.

The successful operation of two-phase Ar avalanche detectors with thin- and thick-GEM multipliers resulted in low detection thresholds, of 4 and 20 primary electrons respectively.

It was found that the double-THGEM multiplier operated in two-phase Ar had slower anode signals as compared to that of the triple-GEM. This can be explained by relatively large ion drift-time between the THGEM electrodes. This slow characteristic signal in the THGEM was demonstrated to be an effective means for rejecting non-avalanche signals when performing off-line pulse-shape analysis.

Noise-rates were assessed with both multipliers in two-phase Ar avalanche mode. At a detection threshold of 4 primary electrons the noise rate of the GEM multiplier was about 0.2 Hz per 1 cm$^2$ of the detector's active area. At a threshold of 20 primary electrons and with pulse-shape analysis, the noise rate of the THGEM multiplier was as low as 0.007 Hz per 1 cm$^2$. These results pave the way towards a new generation of "noiseless" detectors, with noise rates below $10^{-3}$ Hz per kg, as requested in coherent neutrino scattering and other rare-event experiments.

Our general conclusion is that THGEM multipliers can efficiently replace thin-GEMs in two-phase avalanche detectors; they are robust, stable even in cryogenic conditions, can be economically manufactured, have the right performances, including the efficient noise reduction with pulse-shape analysis.

## Acknowledgements

This work was partly supported by the Russian Foundation for Basic Research, by the Israel Science Foundation, grant No 402/05, and by the MINERVA Foundation. A. Breskin is the W.P. Reuther Professor of Research in The Peaceful Use of Atomic Energy.